\documentclass[letterpaper, 10 pt, conference]{ieeeconf}  

\usepackage{graphicx}
\usepackage{amsmath,amssymb,amsfonts}
\usepackage{multirow}
\usepackage{algorithm}
\usepackage{algorithmic}
\usepackage{psfrag}
\usepackage{subfigure}
\usepackage{color}
\usepackage{soul}
\usepackage{xcolor}
\usepackage[bookmarks=false]{hyperref}
\usepackage{lipsum}
\usepackage{mathrsfs}
\usepackage{pbox}
\usepackage{flushend}
\usepackage{mathtools}
\usepackage{wrapfig}

\DeclareMathOperator*{\argmaxA}{arg\,max}

\makeatletter
\def\hlinewd#1{%
  \noalign{\ifnum0=`}\fi\hrule \@height #1 \futurelet
   \reserved@a\@xhline}
\makeatother

\graphicspath{{Figures/}}

\IEEEoverridecommandlockouts                              

\title{\LARGE \bf
RLPG: Reinforcement Learning Approach for Dynamic Intra-Platoon Gap Adaptation for Highway On-Ramp Merging
}

\author{Sushma Reddy Yadavalli$^{1*}$, Lokesh Chandra Das$^{1*}$ and Myounggyu Won$^{1}$
\thanks{$^{*}$Sushma Reddy Yadavalli and Lokesh Chandra Das have equally contributed to this work.}
\thanks{$^{1}$Sushma Reddy Yadavalli, Lokesh Chandra Das and Myounggyu Won are with the Department of Computer Science, University of Memphis, Memphis, TN, United States
        {\tt\small \{sydvalli, ldas, mwon\}@memphis.edu}}%
}

\begin{document}

\maketitle
\thispagestyle{empty}
\pagestyle{empty}

\begin{abstract}
A platoon refers to a group of vehicles traveling together in very close proximity using automated driving technology. Owing to its immense capacity to improve fuel efficiency, driving safety, and driver comfort, platooning technology has garnered substantial attention from the autonomous vehicle research community. Although highly advantageous, recent research has uncovered that an excessively small intra-platoon gap can impede traffic flow during highway on-ramp merging. While existing control-based methods allow for adaptation of the intra-platoon gap to improve traffic flow, making an optimal control decision under the complex dynamics of traffic conditions remains a challenge due to the massive computational complexity. In this paper, we present the design, implementation, and evaluation of a novel reinforcement learning framework that adaptively adjusts the intra-platoon gap of an individual platoon member to maximize traffic flow in response to dynamically changing, complex traffic conditions for highway on-ramp merging. The framework's state space has been meticulously designed in consultation with the transportation literature to take into account critical traffic parameters that bear direct relevance to merging efficiency. An intra-platoon gap decision making method based on the deep deterministic policy gradient algorithm is created to incorporate the continuous action space to ensure precise and continuous adaptation of the intra-platoon gap. An extensive simulation study demonstrates the effectiveness of the reinforcement learning-based approach for significantly improving traffic flow in various highway on-ramp merging scenarios.
\end{abstract}


\section{Introduction}
\label{sec:introduction}

Rapid advances in autonomous vehicles and vehicle-to-everything (V2X) communication technologies are expected to transform the future of transportation systems~\cite{van2018autonomous,chen2017vehicle}. One of the most promising autonomous vehicle applications is a platoon which refers to a group of vehicles driving together in close proximity like a train~\cite{zhou2022cooperative,xu2022truck}. The platooning technology is anticipated to provide numerous benefits to our society~\cite{bhoopalam2018planning}. Traffic efficiency will be enhanced through better utilization of the road space since multiple vehicles drive as a single unit with an extremely small inter-vehicle gap~\cite{jin2020analysis}. The leader-follower driving pattern implemented via autonomous vehicle control will improve driving safety and comfort. Furthermore, the extremely small inter-vehicle gap will reduce aerodynamic drag on vehicles to save fuel consumption~\cite{li2016stabilizing}. 

Although the platooning technology promises to bring significant advantages, complex interactions between a platoon and surrounding vehicles especially their implications on traffic efficiency are not fully understood~\cite{faber2020evaluating}. In particular, we focus on the potential impact of a platoon on the traffic efficiency for highway on-ramp merging~\cite{gao2021optimal,scholte2022control}. More specifically, we aim to mitigate the detrimental effect of the extremely small intra-platoon gap that can make it very difficult for merging vehicles to change lanes to merge into the mainline, noting that a failure to timely merge into the mainline can create a significant speed difference between the mainline and merging traffic, potentially leading to traffic breakdown and even accidents~\cite{xue2022platoon}. 

Simulation studies were conducted to analyze the effect of a platoon on merging traffic on a highway with a ramp~\cite{aramrattana2020simulation, faber2020evaluating,wang2019benefits}. While these studies successfully uncovered the negative effect of a platoon on traffic efficiency, a solution to mitigate the problem was not addressed. Numerous works were proposed to improve traffic efficiency under the presence of a platoon. In particular, state-of-the-art methods rely on control strategies~\cite{gao2021optimal,scholte2022control,fang2022ramp,zhang2022hybrid,xue2022platoon}. Some control-based approaches aim to find the optimal trajectory of a merging vehicle to enhance traffic flow~\cite{gao2021optimal,scholte2022control}. Yet, dynamic adaptation of the intra-platoon gap was not considered. Some solutions are based on a model predictive controller (MPC) to determine the optimal merging gap~\cite{zhang2022hybrid,xue2022platoon}. However, it is extremely challenging for these control-based approaches to effectively account for the complex dynamics of traffic conditions due to the huge computational complexity. 

\begin{figure*}[t]
	\centering
	\includegraphics[width=.99\textwidth]{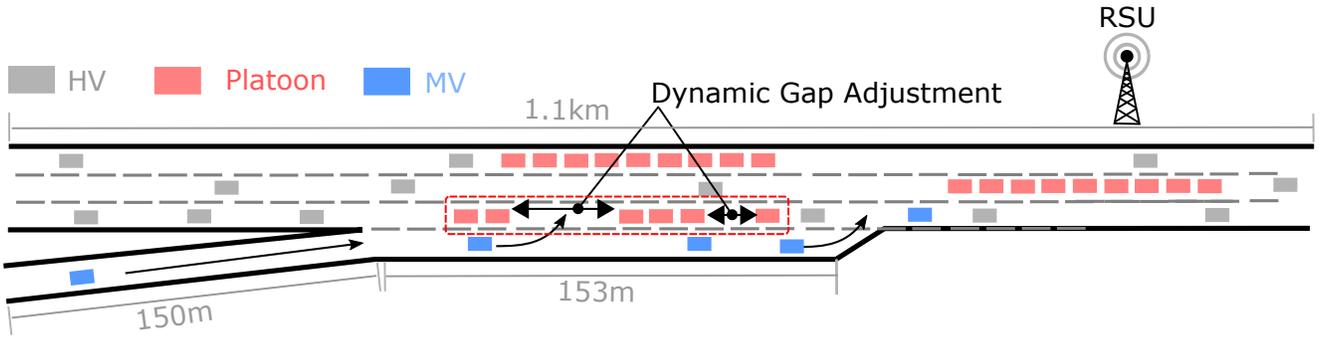}
	\caption {A typical three-lane highway segment with an on-ramp considered for the simulation study. The platoon members of the platoon in the lane adjacent to the merging lane are advised to adjust their intra-platoon gaps individually to allow merging vehicles to change lanes smoothly, thereby improving the overall traffic flow.}
	\label{fig:overview}
\end{figure*}

In this paper, we present RLPG: a reinforcement learning (RL) approach for dynamic intra-platoon gap adaptation for highway on-ramp merging. To the best of our knowledge, RLPG is the first data-driven approach that effectively models the complex dynamics of traffic environments based on RL to enable each platoon member to individually adapt its intra-platoon gap continuously in response to dynamically changing traffic conditions aiming to optimize traffic flow. More specifically, the problem of dynamic intra-platoon gap adaptation for maximizing traffic flow under complex traffic conditions is formulated as a Markov Decision Problem (MDP). A novel RL framework is designed to solve the problem: the state space is created based on a careful review of the transportation literature to incorporate traffic parameters that have a significant impact on the merging efficiency. To support frequent and precise control of the intra-platoon gap, a continuous action space is designed, and an efficient intra-platoon gap decision making algorithm is created based on the deep deterministic policy gradient (DDPG) algorithm to produce an optimal action policy~\cite{lillicrap2015continuous}. Extensive simulations were conducted to train and test the RL model in various highway on-ramp merging scenarios. The results demonstrate the effectiveness of the proposed machine learning-driven approach to improve traffic efficiency under the presence of platoons on a highway with an on-ramp.

The following summarizes the contributions of this paper.

\begin{itemize}
	\item To the best of our knowledge, the proposed solution is the first machine learning-driven approach for adaptively adjusting the intra-platoon gap for individual platoon members in response to dynamically changing traffic conditions.
	\item The RL model is carefully designed in consultation with the latest transportation literature to incorporate critical traffic parameters affecting vehicle merging efficiency. The RL model also incorporates a continuous action space based on the DDPG algorithm to support frequent adjustment of the intra-platoon gap to maximize performance.
	\item Extensive simulations were conducted to evaluate the performance of RLPG under various highway on-ramp merging scenarios with different merging traffic densities, lengths of merging lane, and aggressiveness of merging behavior. 
\end{itemize}


This paper is organized as follows. In Section~\ref{sec:related_work}, we review the latest experimental studies on the effect of platoons for highway on-ramp merging as well as control-based intra-platoon gap adjustment methods. In Section~\ref{sec:motivation}, we conduct a motivational study to demonstrate the significance of a dynamic intra-platoon gap adjustment method for highway on-ramp merging. In Section~\ref{sec:proposed_approach}, we present the details of the proposed method, followed by Section~\ref{sec:results} where we present the simulation results. We then conclude in Section~\ref{sec:conclusion}.

\section{Related Work}
\label{sec:related_work}

Simulation studies were conducted to analyze the impact of platooning focusing on how the small intra-platoon gap affects the traffic efficiency for highway on-ramp merging~\cite{faber2020evaluating,wang2019benefits}. Faber \emph{et al.}  performed a case study of a merging scenario in the presence of a platoon on the A15 motorway in Netherlands~\cite{faber2020evaluating}. Wang \emph{et al.} carried out a simulation to evaluate the effectiveness of different methods for accommodating merging traffic such as courtesy lane change, active yielding, and providing a larger intra-platoon gap~\cite{wang2019benefits}. While increasing the intra-platoon gap alleviated the adverse effect of platooning at merging areas, only a fixed intra-platoon gap was considered, resulting in a limited performance gain.

Aramrattana \emph{et al.} proposed an approach based on varying intra-platoon gaps, addressing the limitation of the fixed intra-platoon gap-based methods~\cite{aramrattana2020simulation}. However, various traffic parameters that can affect the merging efficiency such as the vehicle density, platoon length, vehicle speed, \emph{etc.} were not taken into account in determining the appropriate intra-platoon gap. Karbalaieali \emph{et al.} developed an adaptive algorithm for automated merging control~\cite{karbalaieali2019dynamic}. However, their work is based on a simplifying assumption that one vehicle merges at a time, and a complex traffic condition is simplified into several scenarios, failing to fully account for the dynamics of traffic conditions.

A recent research direction for improving traffic efficiency under the presence of platoons for highway on-ramp merging is focused on control-based approaches~\cite{gao2021optimal,scholte2022control,fang2022ramp,zhang2022hybrid,xue2022platoon}. Gao \emph{et al.} designed a trajectory optimization method for cooperative operation between a platoon and merging vehicles~\cite{gao2021optimal}. However, this approach is concentrated on trajectory planning for a merging vehicle rather than dynamic adaptation of the intra-platoon gap. Scholte \emph{et al.} proposed a novel control strategy based on MPC for intra-platoon gap opening~\cite{scholte2022control}. Nonetheless, similar to~\cite{gao2021optimal}, the control strategy is designed specifically for a single automated vehicle attempting to merge into the mainline. Fang~\emph{et al.} took into account the delay effect of the vehicle-to-infrastructure (V2I) communication in the control method~\cite{fang2022ramp}. Zhang \emph{et al.} developed an MPC-based mixed integer programming model~\cite{zhang2022hybrid} to determine the optimal timing for merging and intra-platoon spacing to accommodate merging vehicles effectively. Xue \emph{et al.} presented a novel control strategy based on a gray prediction model to predict the motion of mainline vehicles and estimate the merging gaps for merging vehicles~\cite{xue2022platoon}. While these static model-based approaches exhibit outstanding performance in certain traffic environments, the generalizability of the control-based solutions in complex and dynamically changing traffic conditions is largely unexplored. 

In contrast to existing control-based approaches, the proposed machine learning-driven approach based on reinforcement learning aims to model the dynamics of complex traffic conditions more effectively to allow each platoon member to individually and adaptively adjust its intra-platoon gap to maximize traffic efficiency.

\section{Motivation: Impact of Platoons on Traffic Flow}
\label{sec:motivation}

We perform a simulation study to analyze the effect of a platoon on traffic efficiency for highway on-ramp merging scenarios. To implement the simulation scenario, we used the traffic simulator SUMO~\cite{lopez2018microscopic}. Various platooning-related functions such as platoon formation, intra-platoon distance control, and configuration of the driving behavior of platoon members were developed using a plugin for SUMO called Simpla~\cite{simpla}. Fig.~\ref{fig:overview} depicts a typical highway on-ramp merging scenario considered in this simulation. The length of the highway segment was 1.1km which had an on-ramp consisting of a 150m merging lane and a 153m acceleration lane. The mainline and merging traffic were generated at a rate of 3600 veh/hour/lane and 1200 veh/hour/lane, respectively. An ``interrupting'' platoon was injected into the adjacent lane to the merging lane in such a way that the platoon arrives at the vehicle-merging area after approximately 30 seconds. An additional platoon was also injected into the other two lanes at a random time. 

\begin{figure}[h]
	\centering
	\includegraphics[width=.9\columnwidth]{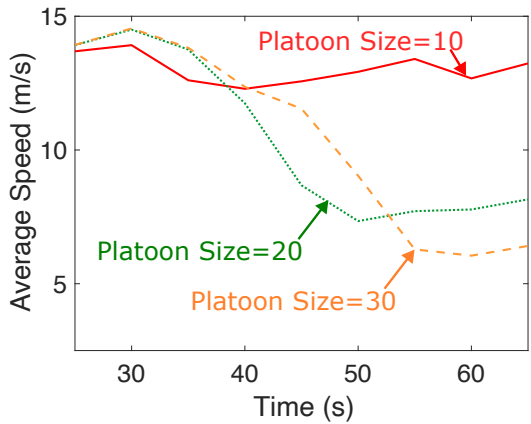}
	\caption {The effect of a platoon on traffic flow with varying platoon sizes. The highway segment could handle platoons with appropriate sizes without impacting the traffic flow. However, the traffic flow started to drop significantly as the platoon size increased.}
	\label{fig:effect_of_platoon_on_traffic_flow}
\end{figure}

To demonstrate the effect of platoons with different sizes on traffic flow, we measured the average speed of the vehicles that passed the highway segment by varying the platoon size. Fig.~\ref{fig:effect_of_platoon_on_traffic_flow} depicts the results. An interesting observation was that small platoons did not cause a noticeable impact on the traffic flow in the simulation scenario. More specifically, it was observed that although the average speed decreased temporarily when the platoon arrived at the merging area, the average speed gradually bounced back up after the platoon passed the merging area. However, such ``resiliency'' was not achievable with larger platoons. We found that the large platoons causing prolonged interruption to merging vehicles led to a traffic breakdown that lasted for a much longer period of time, significantly degrading the overall average speed. More specifically, when the platoon size was 20, the traffic flow decreased by up to 47\%, and in the case of the platoon size 30, traffic flow further decreased by up to 56\%. The results indicate that there exists an optimal platoon size that is large enough yet not causing the traffic breakdown, motivating us to develop a dynamic platoon gap adjustment method to control the ``intra-platoon'' gap effectively to reconfigure the large platoon into smaller ones with the optimal size and maintain the appropriate ``inter-platoon'' gap between them.

\section{RLPG: Reinforcement Learning Approach for Dynamic Intra-Platoon Gap Adaptation}
\label{sec:proposed_approach}

In this section, we present an overview of RLPG (Section~\ref{sec:overview}). We then explain the details of the problem formulation (Section~\ref{sec:problem_formulation}), the RL model design (Section~\ref{sec:rl_design}), and the DDPG algorithm to determine the optimal control policy (Section~\ref{sec:ddpg}). 

\subsection{Overview}
\label{sec:overview}

\begin{figure}[h]
	\centering
	\includegraphics[width=.99\columnwidth]{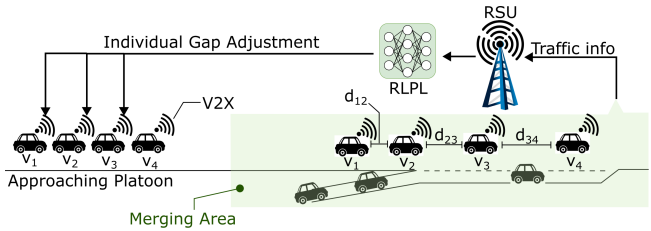}
	\caption {An overview of the operation of RLPG. RLPG computes the optimal individual intra-platoon gap ($d_{12}, d_{23}, d_{34}$) for an approaching platoon consisting of four platoon members $v_1, v_2, v_3$, and $v_4$, based on the current traffic condition. }
	\label{fig:overview2}
\end{figure}

In this section, we explain the overall operation of RLPG for highway on-ramp merging scenarios. Fig.~\ref{fig:overview2} illustrates an overview of the operation of RLPG. As shown, we assume that there exists a roadside unit (RSU) assigned to manage a highway segment. This is a common assumption for many intelligent transportation applications~\cite{wu2012cost,abdrabou2010probabilistic}. An RSU is a platform where RLPG runs. An RSU collects the traffic information from a merging area. In particular, we identify a specific set of traffic parameters that are known in the transportation literature to affect the merging efficiency and allow the RSU to collect such data. The collected traffic information constituting the state space of the RL model is used to determine the optimal intra-vehicle gap for a platoon approaching the merging area. More specifically, the platoon leader of the approaching platoon sends a request containing the current location, size, and speed of the platoon to the RSU to obtain the intra-platoon gap values for its platoon members. The RSU then runs the RLPG model to compute an intra-platoon gap for an individual platoon member and sends the gap values to the platoon leader. Upon receiving the information, the platoon leader broadcasts the optimal values of the intra-platoon gap to its platoon members. Subsequently, each platoon member adjusts its acceleration/deceleration to create the advised intra-platoon gap. The platoon leader keeps communicating with the RSU to receive updated gap control information to effectively respond to dynamically changing traffic conditions.

\subsection{Problem Formulation}
\label{sec:problem_formulation}

We formulate the problem of determining the optimal intra-platoon gap for each platoon member in response to dynamically changing traffic conditions as a Markov Decision Process (MDP)~\cite{sutton2018reinforcement}. MDP is represented as a 4-tuple $M=<S,A,P_{sa},R>$. $S$ is the state space that represents the current traffic condition. $A$ is the action space which is given as a set of intra-platoon gaps for the platoon members. $P_{sa}$ is the probability of having a specific next state assuming that an action $a \in A$ is performed under the current state $s \in S$. $R$ is the reward function that determines a reward $r_t$ for a performed action $a_t$ at time step $t$ under the current state $s_t$. 

We propose to solve the proposed MDP using the reinforcement learning (RL) approach. RL is a widely adopted machine learning method that maps the state space to the action space such that the cumulative sum of the expected reward is maximized. More formally, the cumulative sum of the expected reward 
under a policy $\pi: S \rightarrow A$ that maps the state space to the action space can be expressed as follows.

\begin{equation}
	Q_{\pi}(s,a) = \mathbb{E}_{\pi}[\sum_{k=0}^{\infty}\gamma^kr_{t+k}|s,a],
\end{equation} 

\noindent where $\gamma \in (0,1]$ is a discount factor. Our goal is to find an optimal policy $\pi^*$ such that the cumulative sum of the expected reward is maximized, \emph{i.e.},

\begin{equation}
	Q_{\pi}^*(s,a) = \argmaxA\limits_{\pi} Q_{\pi}(s,a).
\end{equation} 

\subsection{Reinforcement Learning Model}
\label{sec:rl_design}

In this section, we present the details of our RL model. 

\textbf{State space:} A key challenge for designing an effective state space is to identify critical traffic information that is directly related to the vehicle-merging efficiency. We review the transportation literature carefully~\cite{rios2016survey,liao2021cooperative,daamen2010empirical,el2021novel} and incorporate the following traffic information in the state space $S$: the traffic density of both the mainline and ramp~\cite{munjal1971propagation}, the average vehicle speed of both the mainline and ramp~\cite{daamen2010empirical}, the platoon length~\cite{wang2019benefits}, and the intra-platoon gap of each platoon member. 

\textbf{Action space:} The action space $A$ is designed as a range of intra-platoon gap values. The action space is continuous to more effectively control the motion of platoon members. More specifically, we provide a sufficiently large range for the intra-platoon gap (\emph{i.e.,} [2m, 30m]) to allow more than one vehicle to merge in the gap. However, it is worth mentioning that the range can be adapted depending on different road traffic policies. 

\begin{wrapfigure}{r}{0.5\columnwidth}
	\vspace{-12pt}
	\begin{center}
		\includegraphics[width=\linewidth]{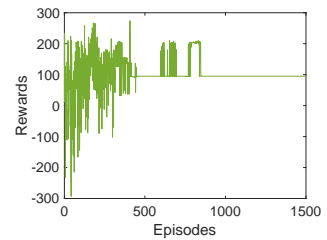}
		\caption {Convergence of the reward function $R$.}
		\label{fig:convergence}
	\end{center}
	\vspace{-10pt}
\end{wrapfigure}

\textbf{Reward function:} Since the main goal of our RL model is to maximize the traffic flow, the reward function $R$ is designed to provide a higher reward if the traffic flow is improved. More precisely, the reward function provides a positive reward if the current average delay is smaller than or equal to the delay for congested traffic conditions denoted by $l_{segment} / v_{congestion}$ where $l_{segment}$ and $v_{congestion}$ are the length of the highway segment and the vehicle speed under the congested condition, respectively. On the other hand, a negative reward is provided if the current average delay is greater than $l_{segment} / v_{congestion}$. Therefore, the reward function $R$ is defined as follows.

\begin{equation}
	R =
	\begin{cases}
		+1 & \mbox{if average delay} \leq \frac{l_{segment}}{v_{congestion}} \\
		-1 & \mbox{if average delay} > \frac{l_{segment}}{v_{congestion}}
	\end{cases}
\end{equation}

With the simulation setting explained in Section~\ref{sec:motivation}, we observed reliable convergence of the reward function as depicted in Fig.~\ref{fig:convergence}.


\subsection{Deep Deterministic Policy Gradient}
\label{sec:ddpg}

A key property of RLPG is the dynamic adaption of the intra-platoon gap in response to time-varying traffic conditions. To support the continuous adjustment of the intra-platoon gap to cope better with dynamically changing traffic environments, we adopt the deep deterministic policy gradient (DDPG) algorithm~\cite{lillicrap2015continuous}, which is highly efficient in finding the optimal policy for a continuous action space, to implement the neural network of our RL model. 

The neural network of our RL model comprises an actor network parameterized by $\theta^\mu$ and a critic network parameterized by $\theta^Q$ to represent the policy $\mu(s|\theta^\mu)$ and the value function $Q(s, a|\theta^Q)$, respectively. In the actor-critic architecture, the actor produces action in terms of the intra-platoon gap given the current state. In particular, the actor explores the environment more effectively by adding noise sampled from the noise process $\mathcal{N}$ to its current action, \emph{i.e.,} ${\mu'}(s_t) = \mu(s_t|\theta_t^\mu) + \mathcal{N}$. The critic network evaluates an action generated by the actor in terms of its impact on the traffic flow. In particular, a reply buffer is used to store the transition ($<s_t, a_t, r_t, s_{t+1}>$) sampled from the environment according to the exploration policy. 

To achieve learning stability of the agent, an actor target network  ${\mu'}(s|\theta^{\mu'})$ and a critic target network, ${Q'}(s, a|\theta^{Q'})$ are incorporated. The target networks are the exact copies of the original networks except that their weights get a slow update unlike the online networks. In other words, the weights of the target networks are updated based on ``soft'' target updates rather than directly copying the main network weights. The following equation represents the weight update process for the actor and critic target networks.

\begin{equation}\label{soft_update}
	\begin{split}
		\theta^{Q'} \leftarrow \tau\theta^Q + (1-\tau)\theta^{Q'},\\
		\theta^{\mu'} \leftarrow \tau\theta^\mu + (1-\tau)\theta^{\mu'},
	\end{split}
\end{equation}

\noindent where $\tau \ll 1$ is an update parameter.

The actor network updates its policy using the sampled policy gradient method and adjusts its parameters according to the following equation to produce an optimal action.

\begin{equation}\label{actor_policy_update}
	\nabla_{\theta^\mu}J \approx \frac{1}{N} \sum_{i}\nabla_a Q(s, a|\theta^Q)|_{s=s_i, a=\mu(s_i)}\nabla_{\theta^\mu}\mu(s|\theta^\mu)|_{s_i}.
\end{equation}

\begin{algorithm}[t]
	\begin{algorithmic}[1]
		\REQUIRE {The maximum episodes $M$, time range for each episode $T$, discount factor $\beta$, updating rate $\tau$, reward function $R$.}
		\ENSURE{The intra-platoon gap policy.}
		\STATE Initialize the critic network $Q(s, a|\theta^Q)$ and the actor network $\mu(s|\theta^\mu)$ with random weights $\theta^Q$ and $\theta^\mu$.
		\STATE Initialize the target network parameters $Q'$ and $\mu'$ as well as weights $\theta^{Q'} \leftarrow \theta^Q$, $\theta^{\mu'} \leftarrow \theta^\mu$ for the critic and actor networks, respectively.
		\STATE Initialize the reply buffer \textbf{B}.
		\FOR {$episode = 1 : M$}
		\STATE Initialize a noise process $\mathcal{N}$ for action exploration.
		\STATE Get an initial observation $s_1$.
		\FOR {$t = 1 : T$}
		\STATE Select action $a_t = \mu(s_t|\theta^\mu)+\mathcal{N}_t$ with respect to the current policy and the random exploration noise.
		\STATE Execute action $a_t$, and observe reward $r_t$ and new observation $s_{t+1}$.
		\STATE Store the transition $(s_t, a_t, r_t, s_{t+1})$ in \textbf{B}.
		\STATE Sample a mini-batch of N transitions $(s_i, a_i, r_i, s_{i+1})$ randomly from \textbf{B} for the main network.
		\STATE Evaluate $y_i = r_i + \beta Q'(s_{i+1}, {\mu'}(s_{i+1}|\theta^{\mu'})|\theta^{Q'})$.
		\STATE Update the critic network by minimizing the loss in the main network: $L=\frac{1}{N}\Sigma_{i}(y_i - Q(s_i, a_i|\theta^Q))^2$.
		\STATE Update the actor network policy using the sampled policy gradient in the main network: $\nabla_{\theta^\mu}J \approx \frac{1}{N} \sum_{i}\nabla_a Q(s, a|\theta^Q)|_{s=s_i, a=\mu(s_i)}\nabla_{\theta^\mu}\mu(s|\theta^\mu)|_{s_i}$.
		\STATE Update the target network using the following: $$\theta^{Q'} \leftarrow \tau\theta^Q + (1-\tau)\theta^{Q'}$$ $$\theta^{\mu'} \leftarrow \tau\theta^\mu + (1-\tau)\theta^{\mu'}$$
		\ENDFOR
		\ENDFOR
	\end{algorithmic}
	\caption{The DDPG Algorithm for Intra-Platoon Gap Decision Making}
	\label{DDPG_Algorithm}
\end{algorithm}

The critic network uses the Bellman equation to update its value function given the actor policy and optimizes its loss function using the following equations.

\begin{equation}\label{critic_loss}
	\begin{split}
		L=\frac{1}{N}\Sigma_{i}(Q(s_i, a_i|\theta^Q) -y_i)^2,\\
		y_i = r_i + \beta Q'(s_{i+1}, \mu'(s_{i+1}|\theta^{\mu'})|\theta^{Q'}),
	\end{split}
\end{equation}

\noindent where $N$ is the batch size, $r_i$ is immediate reward at $i^{th}$ time-step, $\beta$ is a discount factor, $Q'(s_{i+1}, \mu'(s_{i+1}|\theta^{\mu'}))$ is the target value for the state-action pair $(s_{i+1}, \mu'(s_{i+1}|\theta^{\mu'})$, and $Q(s_i, a_i)$ is the value learned from the online network. The training process for our RL model based on the DDPG algorithm is outlined in Algorithm~\ref{DDPG_Algorithm}.

\section{Simulation Results}
\label{sec:results}

\subsection{Simulation Setup}
\label{sec:setting}

We implemented RLPG on a traffic simulator called SUMO~\cite{lopez2018microscopic}. In particular, Simpla, a plugin for SUMO, was integrated to utilize platoon-related functionality such as platoon formation, intra-platoon gap control, the configuration of the driving behavior of platoon members, \emph{etc.}~\cite{simpla}. The RL model of RLPG was developed based on Keras and Tensorflow~\cite{abadi2016tensorflow}. The RL model was interfaced with SUMO via Traffic Control Interface (TraCI) to manipulate the simulated objects in an online fashion~\cite{wegener2008traci}. A workstation equipped with the Intel Xeon Gold 5222 Processor, NVIDIA® RTX™ A4000, and 48GB RAM running on Windows 11 OS was used to train and test the RL model. 

We now explain the simulation setup. A highway segment with an on-ramp was created as depicted in Fig.~\ref{fig:overview}. The length of the highway segment was 1.1km. The length of the merging lane of the on-ramp was set to 150m. The ramp had a 150m-long acceleration lane which was used by merging vehicles to accelerate before they merge into the mainline. The traffic for the mainline was generated at a rate of 3,600 veh/hour/lane. The merging traffic was varied for evaluation purposes. Vehicles used in the simulation had a body length of 4$\sim$5m. A widely used lane-change model~\cite{erdmann2015sumo} was adopted to implement the lane-changing behavior of merging vehicles. The average speed of the vehicles that passed the highway segment was used as the main metric to evaluate the traffic flow. 

\begin{table}[h]
	\centering
	\caption{The Optimal Hyperparameter Values for Actor-Critic Neural Network of Proposed RL Model}
	\label{tab:my-table}
		\begin{tabular}{|p{2.5cm}|p{4.5cm}|}
			\hline
			\multicolumn{2}{|c|}{Hyperparameters} \\ \hline \hline
			\multicolumn{1}{|c|}{Actor Learning Rate} & 0.001 \\ \hline
			\multicolumn{1}{|c|}{Critic Learning Rate} & 0.001 \\ \hline
			\multicolumn{1}{|c|}{Discount Factor $\beta$} & 0.99 \\ \hline
			\multicolumn{1}{|c|}{Batch Size} & 32 \\ \hline
			\multicolumn{1}{|c|}{Reply Buffer} & 50k \\ \hline
			\multicolumn{1}{|c|}{Weight Update Rate $\tau$} & 0.001 \\ \hline
			\multicolumn{1}{|c|}{Standard Deviation} & 0.2 \\ \hline
			\multicolumn{1}{|c|}{Actor Hidden Layers} & 2 hidden layers with 32, 16 hidden units, respectively \\ \hline
			\multicolumn{1}{|c|}{Critic Hidden Layers} & 5 hidden layers with 32, 16, 8, 16, 8 hidden units, respectively \\ \hline
			\multicolumn{1}{|c|}{Activation Function} & ReLU for hidden layers; 'tanh' for actor output layer; and 'linear' for critic output layer \\ \hline
		\end{tabular}%
\end{table}
	
To obtain the best results, we performed hyperparameter tuning to find an optimal set of hyperparameters before conducting simulations. More specifically, we used the grid search method to find the optimal values for the actor learning rate, critic learning rate, discount factor, batch size, reply buffer size, weight update rate, and the number of layers for the actor network and critic network. Table 1 summarizes the hyperparameter values identified through the hyperparameter tuning process. The Adam optimizer~\cite{kingma2014adam} was utilized for both the actor and critic networks. For exploration and exploitation, the Ornstein-Uhlenbeck process with a standard deviation of 0.2 was used~\cite{uhlenbeck1930theory}.

\subsection{Effect of Merging Traffic Density}
\label{sec:merging_traffic_density}
	
We evaluate the performance of RLPG by varying several important traffic parameters including the merging traffic density, length of the merging lane, and lane change behavior. First, we analyze how RLPL performs for highway on-ramp merging scenarios with different merging traffic densities. The merging traffic density is defined as the number of merging vehicles on the merging lane. Since RLPG is the first approach for dynamic intra-platoon gap adaptation, the performance of RLPG is compared with the base case where no vehicle is equipped with an intra-platoon gap adjustment mechanism.  

\begin{figure}[h]
	\centering
	\begin{minipage}{.49\columnwidth}
		\centering
		\includegraphics[width=\linewidth]{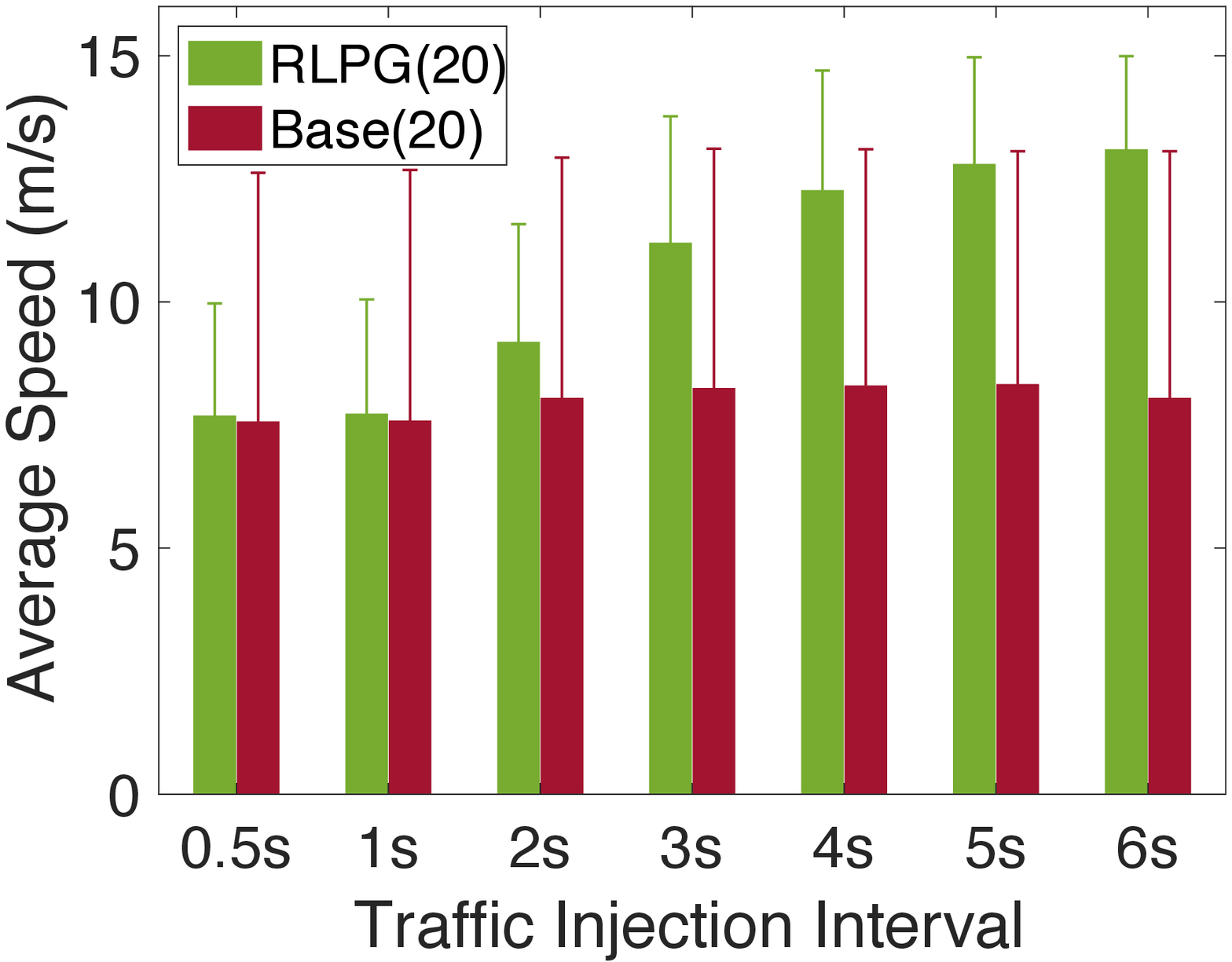}
		\caption{The effect of the merging traffic density on the performance of RLPG for the platoon size of 20.}
		\label{fig:density20}
	\end{minipage}%
	\hspace*{3mm}
	\begin{minipage}{.49\columnwidth}
		\centering
		\includegraphics[width=\linewidth]{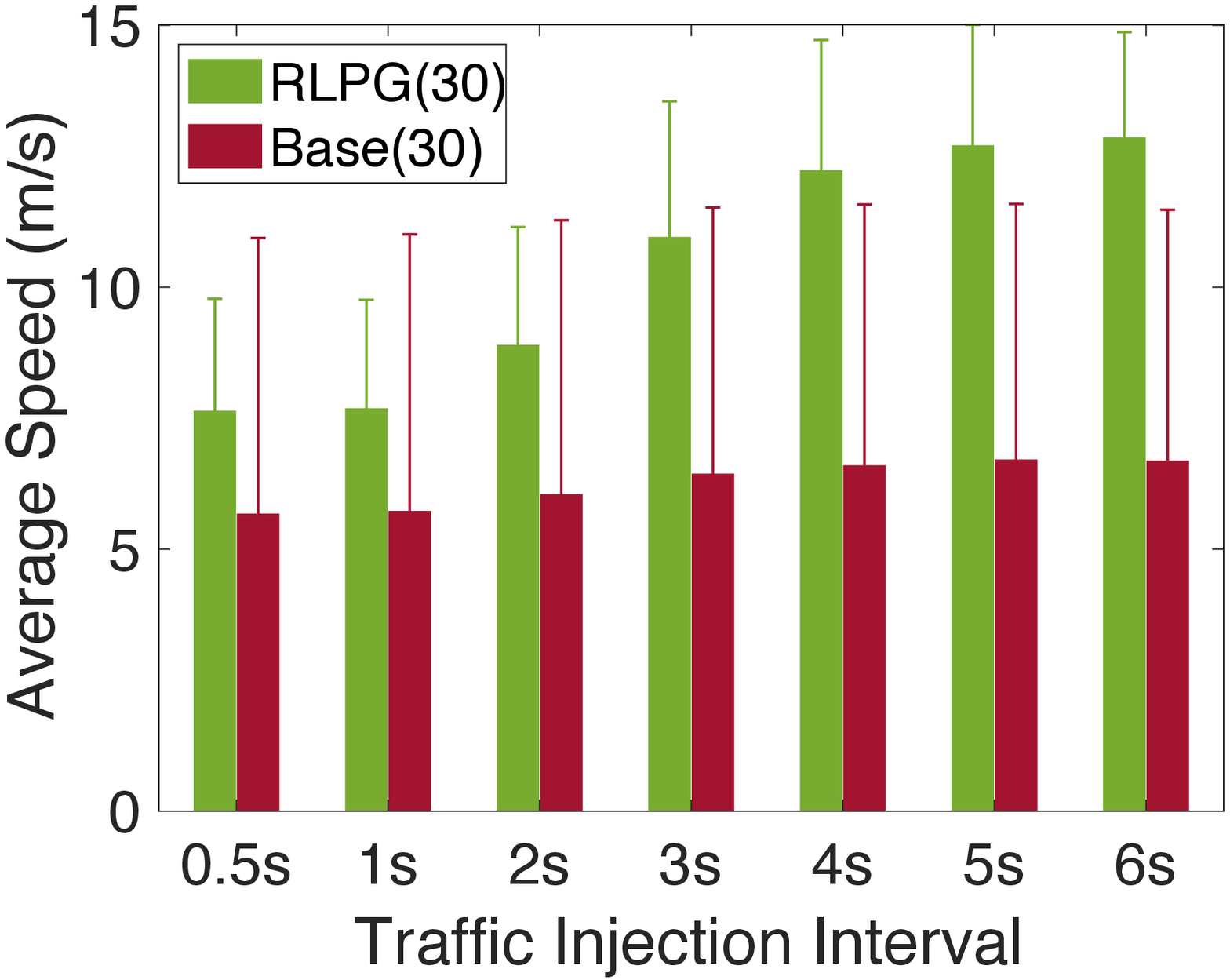}
		\caption{The effect of the merging traffic density on the performance of RLPG for the platoon size of 30.}
		\label{fig:density30}
	\end{minipage}
\end{figure}

Figs.~\ref{fig:density20} and~\ref{fig:density30} display the results for the platoon sizes of 20 and 30, respectively. To control the merging traffic density, we varied the merging traffic injection interval, \emph{i.e.,} a higher traffic injection interval means a lower traffic density and vice versa. The results demonstrate that RLPG improves the traffic flow significantly. The average speed was increased by 31.8\% compared with the base case when the platoon size was 20. The performance gain was even more higher when the platoon size was larger. RLPG achieved higher average speed by 66.3\% compared with the base case when the platoon size was 30.

We also observed that RLPG exhibited noticeably robust performance regardless of the platoon size. More specifically, while the traffic flow decreased by 21.8\% when the platoon size was increased from 20 to 30 for the base case, RLPG successfully suppressed the negative impact of the longer platoon limiting the performance degradation by only 1.3\%. Another interesting observation was the high standard deviation for the base case. Such a high standard deviation represents the instability of the traffic flow caused by the platoon passing the highway segment. On the other hand, RLPG  achieved consistently high traffic flow. 

\begin{figure}[h]
	\centering
	\includegraphics[width=.99\columnwidth]{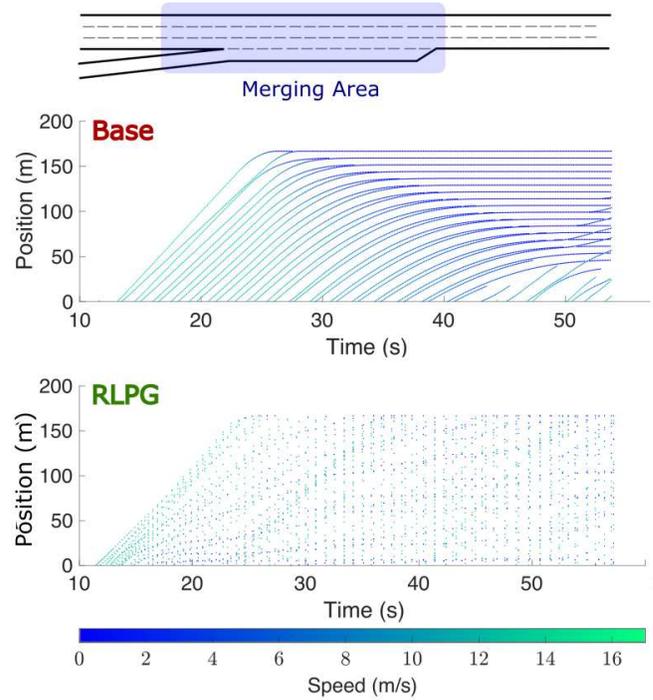}
	\caption {A space-time graph for RLPG and the base case, representing the time-varying traffic flow at horizontal positions over a 200m-long merging area. The result demonstrates a traffic breakdown caused by a passing platoon, which led to an overall degradation of traffic flow. The result also demonstrates that RLPG prevents such a traffic breakdown, thereby maintaining high traffic flow.}
	\label{fig:space}
\end{figure}

A space-time graph depicted in Fig.~\ref{fig:space} provides a better view of how RLPG improves traffic flow. The graph shows the time-varying average speed measured at a certain horizontal position over the 200m merging area. The result for the base case demonstrates that the average speed started to be recorded at around 15s when vehicles on the mainline arrived at the merging area (Note that initially, there is no vehicle in the merging region yet). A traffic breakdown occurred near the end of the merging area at about 25s. The traffic breakdown propagated upstream leading to the degradation of traffic flow over the entire merging area. On the other hand, Fig.~\ref{fig:space} indicates that the adaptive intra-platoon gap adjustment for RLPG successfully prevented the traffic breakdown, thereby sustaining high traffic flow. 

\subsection{Effect of Length of Merging Lane}
\label{sec:length_of_ramp}

In this section, we evaluate the performance of RLPG by varying the length of the merging lane. In general, a longer merging lane allows for smoother lane changes for merging vehicles, potentially leading to better traffic flow~\cite{sarvi2007microsimulation}. Nonetheless, it is not clear if traffic flow will be improved with the longer merging lane when a platoon is present. In particular, we aim to find how RLPG reacts to different lengths of merging lane. 

\begin{figure}[h]
	\centering
	\begin{minipage}{.49\columnwidth}
		\centering
		\includegraphics[width=\linewidth]{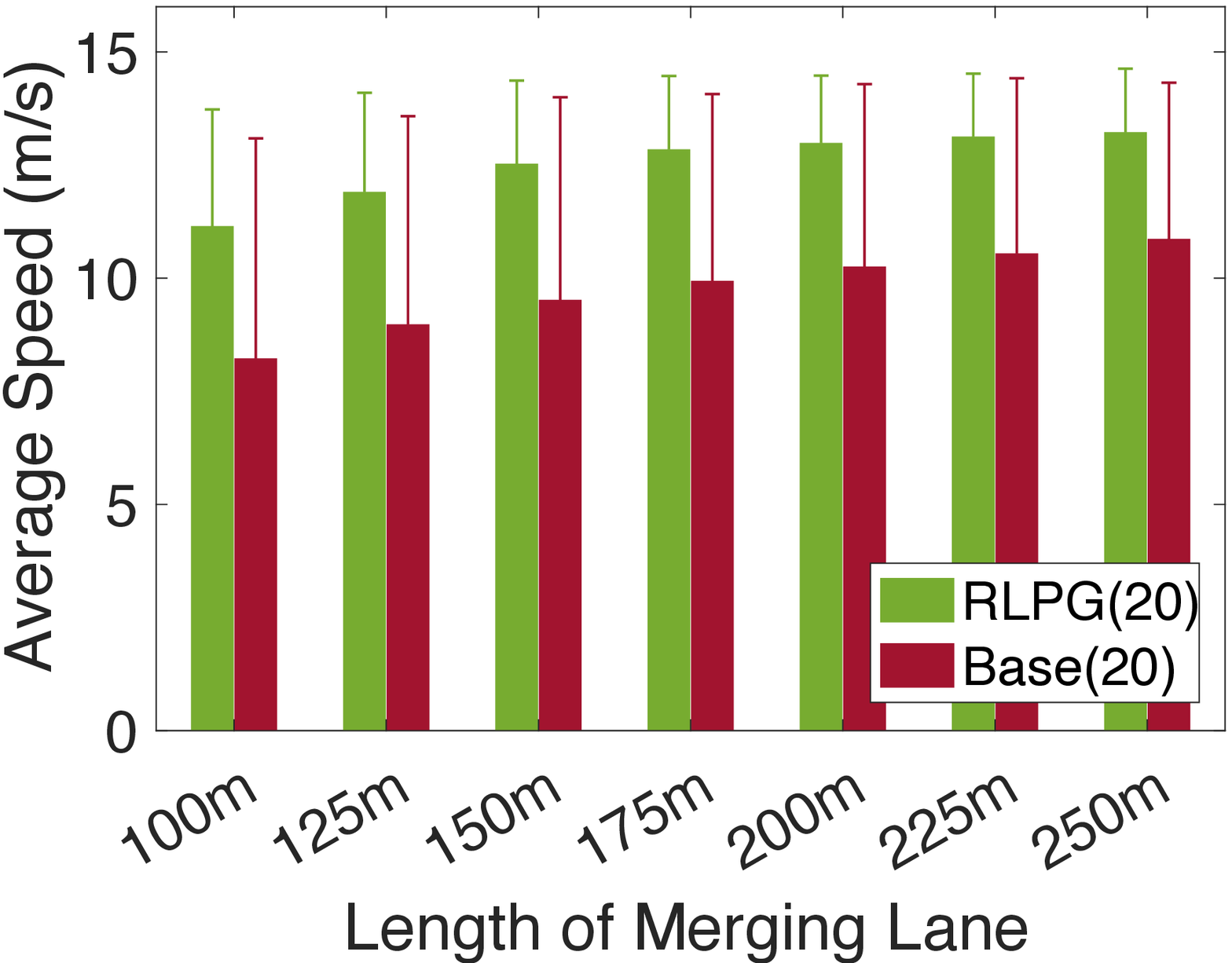}
		\caption{The effect of the length of merging lane on the performance of RLPG for the platoon size of 20.}
		\label{fig:length20}
	\end{minipage}%
	\hspace*{3mm}
	\begin{minipage}{.49\columnwidth}
		\centering
		\includegraphics[width=\linewidth]{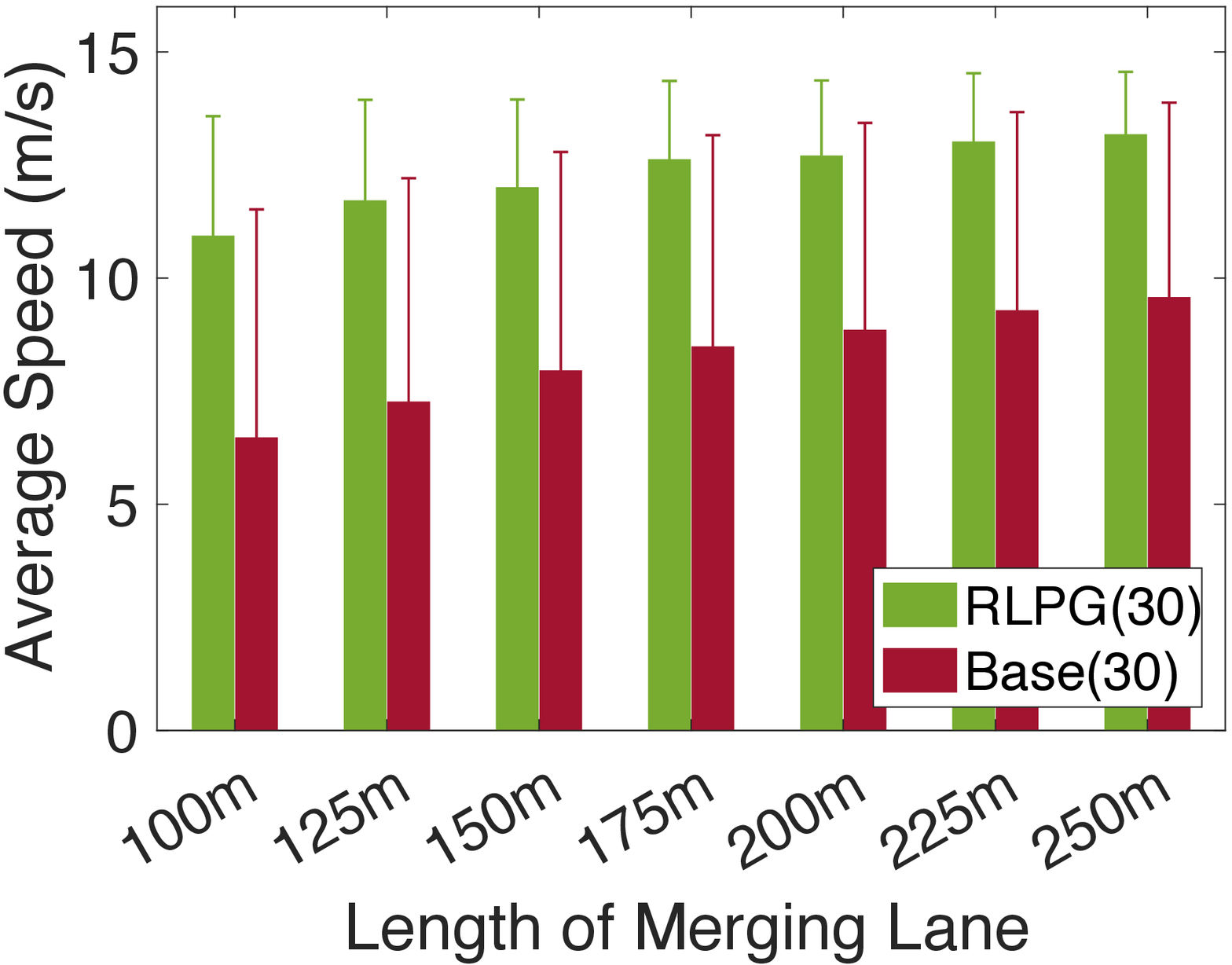}
		\caption{The effect of the length of merging lane on the performance of RLPG for the platoon size of 30.}
		\label{fig:length30}
	\end{minipage}
\end{figure}
	
Figs~\ref{fig:length20} and~\ref{fig:length30} depict the results for the platoon sizes of 20 and 30, respectively. We observed that RLPG  improved traffic flow consistently regardless of the length of merging lane in comparison with the base case. In particular, RLPG improved traffic flow by up to 28.4\% compared with the base case. We also noted that the performance gain for the larger platoon (30) was higher by up to 15.9\% compared with that for the smaller one (20).

Although the results confirm the degradation of traffic flow for shorter merging lanes even for RLPG, an interesting observation was that RLPG effectively suppressed the negative impact of shorter merging lane. More specifically, while the traffic flow was decreased by 24.3\% (platoon size=20) and 32.4\% (platoon size=30) for the base case when the length of the merging lane was decreased from 250m to 100m, the traffic flow was reduced by only 15.7\% (platoon size=20) and 16.9\% (platoon size=30) when RLPG was applied. The result demonstrates the high applicability of the proposed approach for various highway on-ramp merging scenarios with different types of merging lanes.

\subsection{Effect of Lane-Changing Behavior}
\label{sec:lane_changing_behavior}
	
In this section, we evaluate the effect of the lane-changing behavior of merging vehicles on the performance of RLPG. To control the lane-changing behavior, we adopt a widely used lane-changing model~\cite{erdmann2015sumo}. More specifically, in this lane-changing model, a parameter ``lcAssertive'' is used to adjust the willingness of a driver to change lanes. In other words, the parameter represents the aggressiveness of the lane-changing behavior. 
	
\begin{figure}[h]
	\centering
	\begin{minipage}{.49\columnwidth}
		\centering
		\includegraphics[width=\linewidth]{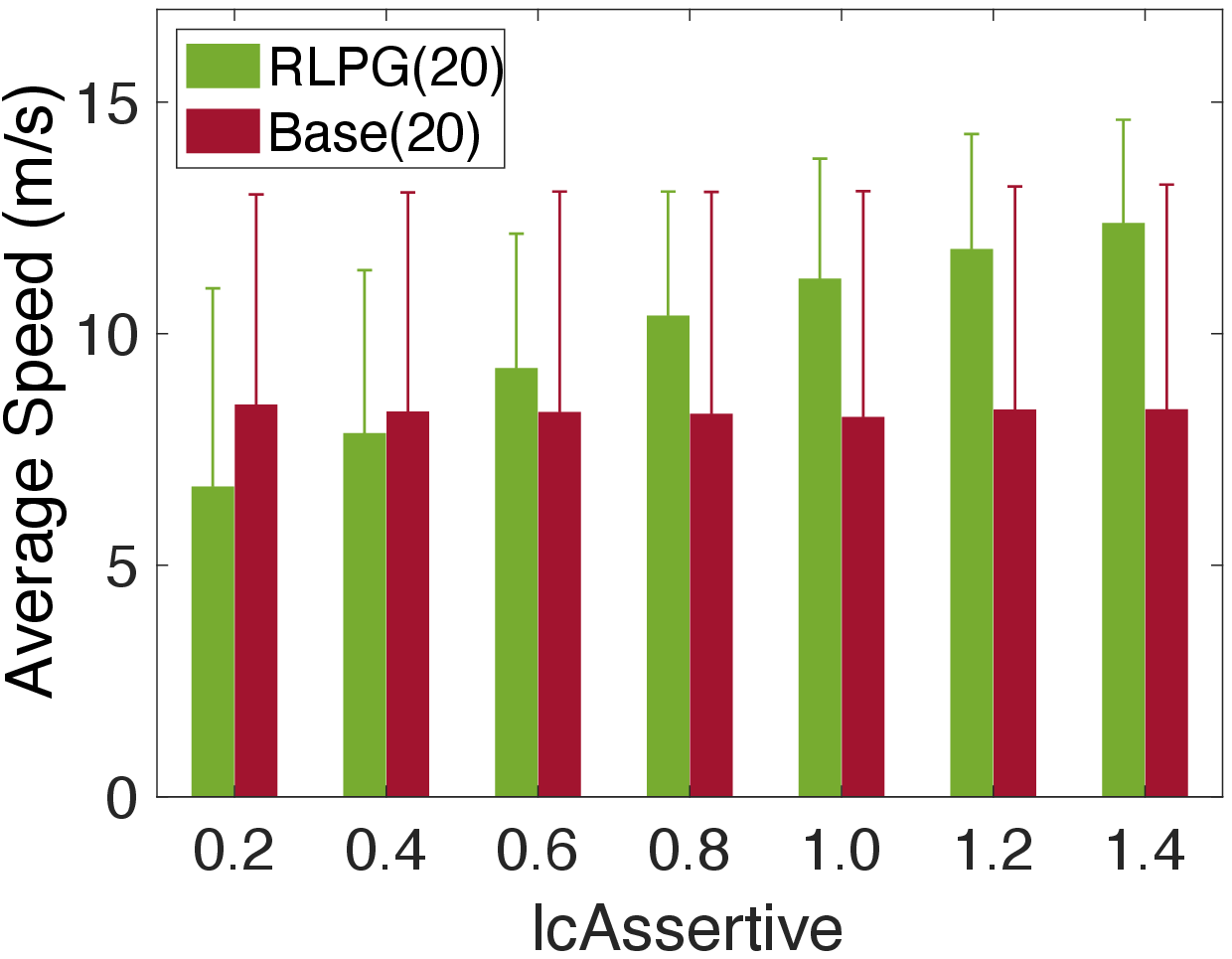}
		\caption{The effect of lane changing behavior on the performance of RLPG for the platoon size of 20.}
		\label{fig:behavior20}
	\end{minipage}%
	\hspace*{3mm}
	\begin{minipage}{.49\columnwidth}
		\centering
		\includegraphics[width=\linewidth]{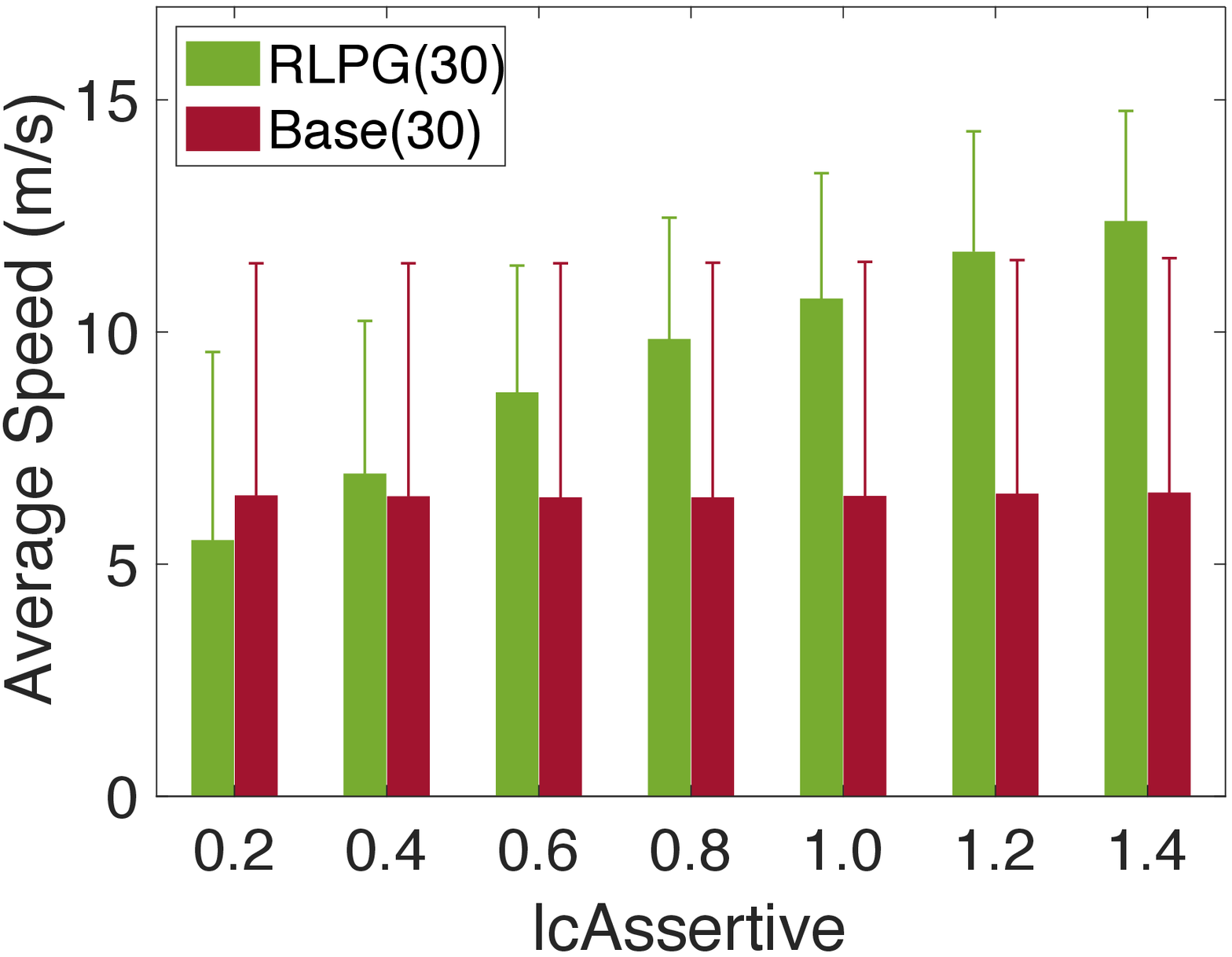}
		\caption{The effect of lane changing behavior on the performance of RLPG for the platoon size of 30.}
		\label{fig:behavior30}
	\end{minipage}
\end{figure}
	
Figs~\ref{fig:behavior20} and~\ref{fig:behavior30} depict the results for the platoon size of 20 and 30, respectively. We observed that the aggressiveness of the lane-changing behavior of merging vehicles did not have much impact on traffic flow for the base case. The reason can be attributed to the fact that the intra-platoon gap was so small that merging vehicles did not attempt to change lanes regardless of the aggressiveness of the lane-changing behavior. In contrast, the effect of the lane-changing behavior was more noticeable for RLPG because RLPG dynamically adjusted the intra-platoon gap so that merging vehicles could merge into the space. Specifically, we observed that, for the platoon size of 20, the traffic flow increased by 19.4\% on average when the lcAssertive value was increased from 0.25 to 1.4 (\emph{i.e.,} more aggressive lane-changing behavior). Additionally, we observed that the effect of aggressive lane-changing behavior on improving traffic flow was even more apparent when the platoon size was larger. More specifically, for the platoon size of 30, the traffic flow was enhanced  by 45.2\% on average as we increased the lcAssertive value from 0.25 to 1.4. 

\subsection{Computational Delay}
\label{sec:computation_time}

	
Due to the fast speed of vehicles and platoons on a highway, it is necessary for the RSU to quickly compute the intra-platoon gap for individual platoon members. Especially, the continuous adaptation of the intra-platoon gap in response to dynamically changing traffic conditions requires even smaller computational delay. In this section, we measure the computation delay to demonstrate the feasibility of the proposed solution.

\begin{wrapfigure}{r}{0.6\columnwidth}
	\vspace{-18pt}
	\begin{center}
		\includegraphics[width=\linewidth]{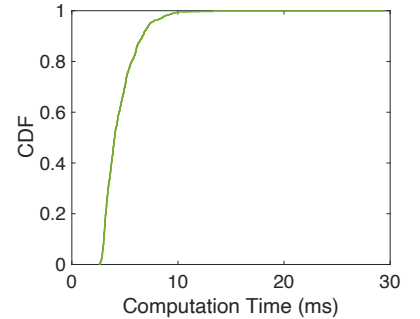}
		\caption {A cumulative distribution function (CDF) graph of the computation delay. The average delay was 4.5ms which is small enough to apply RLPL for highway on-ramp scenarios with fast speed vehicles.}
		\label{fig:computation_time}
	\end{center}
	\vspace{-10pt}
\end{wrapfigure}


We measured the computational delay repeatedly for 1,000 times. Fig.~\ref{fig:computation_time} depicts the cumulative distribution function (CDF) graph of the measured computational delay. The result shows that the computation delay was within 10ms in over 90\% cases. It also indicates that the average computational delay was only 4.5ms. Overall, the computational delay of RLPG was sufficiently small to be applicable to highway on-ramp merging scenarios with high-speed platoons dynamically adapting their intra-platoon gaps.

\section{Conclusion}
\label{sec:conclusion}
	
We have presented RLPG: a reinforcement learning-based approach for adaptive adjustment of the intra-platoon gap to maximize traffic flow for highway on-ramp merging. To the best of our knowledge, RLPL is the first data-driven approach based on RL that effectively captures the dynamics of complex traffic conditions for frequent, precise, and individual adaptation of the intra-platoon gap for each platoon member, addressing the limitations of existing control-based methods. Extensive simulations were conducted to demonstrate that RLPG significantly improves traffic flow under various highway on-ramp merging scenarios with varying traffic densities, merging lane lengths, and lane-changing behaviors. 

The challenge of examining the impact of platoons on traffic efficiency across various road types and developing a highly adaptable RL model for diverse road conditions remains an open problem. Furthermore, while current studies primarily concentrate on improving traffic flow, exploring the simultaneous optimization of traffic efficiency, safety, and driving comfort presents an intriguing avenue for future research. Anticipated as a valuable resource, our proposed RL-based solution is expected to inspire further research endeavors towards the creation of effective machine learning-driven techniques, aimed at expediting the adoption of platooning technology.

\bibliographystyle{IEEEtran}
\bibliography{mybibfile}

\end{document}